%% file: paper.tex
\documentclass[twocolumn,showpacs,aps,prl,superscriptaddress]{revtex4}

\usepackage{graphicx}
\usepackage{dcolumn}
\usepackage{amsmath}
\usepackage{epsfig}

\input babarsym

\newcommand{\PERM}{\ensuremath{\times 10^{-3}}}
\newcommand{\DSS} {\ensuremath{D^{**}}}
\newcommand{\DSSP}{\ensuremath{D^{**+}}}
\newcommand{\DSSZ}{\ensuremath{D^{**0}}}
\newcommand{\DSSL}{\ensuremath{D^{**}\ell}}
\newcommand{\DI}  {\ensuremath{D_1}}
\newcommand{\DII} {\ensuremath{D_2^*}}
\newcommand{\DIZ} {\ensuremath{D_1^0}}
\newcommand{\DIIZ}{\ensuremath{D_2^{*0}}}
\newcommand{\DIP} {\ensuremath{D_1^+}}
\newcommand{\DIM} {\ensuremath{D_1^-}}
\newcommand{\DIIP}{\ensuremath{D_2^{*+}}}
\newcommand{\DIIM}{\ensuremath{D_2^{*-}}}
\newcommand{\DS}  {\ensuremath{D^{*}}}
\newcommand{\DSZ} {\ensuremath{D^{*0}}}
\newcommand{\DSP} {\ensuremath{D^{*+}}}
\newcommand{\DZ}  {\ensuremath{D^0}}
\newcommand{\DP}  {\ensuremath{D^+}}
\newcommand{\D}   {\ensuremath{D}}

\newcommand{\KM}  {\ensuremath{K^-}}
\newcommand{\PIP} {\ensuremath{\pi^+}}
\newcommand{\PIM} {\ensuremath{\pi^-}}
\newcommand{\PIZ} {\ensuremath{\pi^0}}
\newcommand{\PI}  {\ensuremath{\pi}}
\newcommand{\TO}  {\ensuremath{\to}}
\newcommand{\BDSSLN} {\ensuremath{B \to D^{**}\ell\nu}}
\newcommand{\COSBY}  {\ensuremath{cos_{BY}}}
\newcommand{\COSBYP} {\ensuremath{cos_{BY^\prime}}}
\newcommand{\dM}     {\ensuremath{\Delta m}}
\newcommand{\ADI}    {\ensuremath{A_{\DI}}}
\newcommand{\BDII}   {\ensuremath{\BR_{D/D^{(*)}}}}

\newcommand{\BABARPubYear}    {08}
\newcommand{\BABARPubNumber} {043}

\newcommand{\SLACPubNumber}{13348}
\newcommand{\LANLNumber}{0808.333}
\newcommand{\PRLNumber}{{\bf 103,} 051803}

\def\figurebox#1#2#3{%
    \def\arg{#3}%
    \ifx\arg\empty
    {\hfill\vbox{\hsize#2\hrule\hbox to #2{\vrule\hfill\vbox to #1{\hsize#2\vfill}\vrule}\hrule}\hfill}%
    \else
    {\hfill\epsfbox{#3}\hfill}%
    \fi}

\begin{document}

\preprint{\babar-PUB-\BABARPubYear/\BABARPubNumber}
\preprint{SLAC-PUB-\SLACPubNumber} 

\begin{flushleft}
\babar-PUB-\BABARPubYear/\BABARPubNumber\\
SLAC-PUB-\SLACPubNumber\\
arXiv:\LANLNumber\ [hep-ex]\\
PRL \PRLNumber\\
\end{flushleft}

\begin{flushright}
\end{flushright}

\title{
{\large \bf
\vspace*{18mm}
Measurement of Semileptonic {\boldmath $B$} Decays into\\
Orbitally Excited Charmed Mesons} 
}

\input authors_jul2008

\date{August 5, 2008}

\begin{abstract}
We present a study of $B$ decays into semileptonic final states containing
charged and neutral \DI(2420) and \DII(2460).
The analysis is based on a data sample of $208 \invfb$ collected at the
\FourS\ resonance with the \babar\ detector at the \pep2\ asymmetric-energy $B$
Factory at SLAC.
With a simultaneous fit to four different decay chains, the semileptonic branching
fractions are extracted from measurements of the mass difference $\dM=m(\DSS)-m(\D)$
distributions.
Product branching fractions are determined to be
$\BR (\Bp \to \DIZ  \ell^+ \nu_{\ell}) \times \BR(\DIZ\TO\DSP\PIM)
= ( 2.97 \pm 0.17 \pm 0.17)\PERM$,
$\BR (\Bp \to \DIIZ  \ell^+ \nu_{\ell})\times \BR(\DIIZ\TO\D^{(*)}{}^{+}\PIM) 
= ( 2.29 \pm 0.23 \pm 0.21)\PERM$,
$\BR (\Bz \to \DIM  \ell^+ \nu_{\ell}) \times \BR(\DIM\TO\DSZ\PIM)
= ( 2.78 \pm 0.24 \pm 0.25)\PERM$
and 
$\BR (\Bz \to \DIIM  \ell^+ \nu_{\ell})\times \BR(\DIIM\TO\D^{(*)}{}^{0}\PIM)
= ( 1.77 \pm 0.26 \pm 0.11)\PERM$.
In addition we measure the branching ratio
 $\Gamma(\DII\TO\D\PIM)/\Gamma(\DII\TO\D^{(*)}\PIM) = 0.62 \pm 0.03 \pm 0.02$.
\end{abstract}


\maketitle

Measurements of the Cabbibo-Kobayashi-Maskawa matrix elements $|V_{cb}|$ and $|V_{ub}|$ rely on
precise knowledge of semileptonic $B$-meson decays.
Decays with orbitally excited charm mesons (\DSS) in the final state give a
significant contribution to the total semileptonic decay rate.
A better understanding of these decays will reduce the uncertainty in the composition of
the signal and backgrounds for inclusive and exclusive measurements~\cite{B-decay}.

In the framework of Heavy Quark Symmetry (HQS), \DSS\ mesons
form two doublets with $j_q^P=1/2^-$ and $j_q^P=3/2^-$ where $j_q^P$
denotes the spin-parity of the light quark coupled to the orbital angular momentum.
The doublet with $j_q^P=3/2^-$, namely the \DI\ and \DII, have to decay via D wave to
conserve parity and angular momentum and therefore are narrow with widths of order of
10 \mev~\cite{DSS-theory}. 
The relative contribution of the two doublets and the polarization of the produced
\DSS\ mesons can be compared with QCD sum rules \cite{sum-rules} and predictions from
Heavy Quark Effective Theory \cite{LLSW}.

In this Letter we describe a simultaneous measurement of all $B$ semileptonic decays to the
two narrow orbitally-excited charmed states, without explicit reconstruction
of the rest of the event.
The CLEO collaboration has previously reported a branching fraction
measurement for $\Bp\to\DIZ\ell^+\nu$
and an upper limit for $\Bp\to\DIIZ\ell^+\nu$~\cite{DSS-CLEO}. 
Belle and \babar\ have reported results using a technique in
which one of the $B$ mesons in the process $\FourS\to\BB$ is fully
reconstructed~\cite{DSS-recoil}.

In this analysis we use a sample with a total integrated luminosity of 208\invfb,
part of the complete data set collected with the \babar\ detector at the \pep2\ storage ring,
operating at a center of mass energy of 10.58\gev.

The \babar\ detector~\cite{babar-det} and event reconstruction~\cite{babar-reco}
are described in detail elsewhere.
A Monte Carlo (MC) simulation of the detector based on GEANT4~\cite{GEANT} is used to
estimate signal efficiencies and to understand the backgrounds. The sample of simulated
\BB\ events is equivalent to approximately three times the data sample and
a dedicated simulation of signal events based on the ISGW2 model~\cite{ISGW2}
has been produced with statistics equivalent to roughly 5 times the expected signal
yield contained in the data.

\DSS\ decays are reconstructed in the decay chains \DSS\TO\DS\PIM~\cite{conjugate},
and \DSS\TO\D\PIM. The former is accessible to both narrow \DSS\ states while the
latter has no contribution from the \DI.
Intermediate \DS\ states are reconstructed in \DS\TO\DZ\PI\ and the \D\ mesons
are reconstructed exclusively in \DZ\TO\KM\PIP\
and \DP\TO\KM\PIP\PIP. \DSS\ candidates are then paired with reconstructed leptons
and required to be consistent with the semileptonic decays \BDSSLN, as described in
the following.

First, events which are most likely to contain a semileptonic $B$ decay are selected.
We require that there is a reconstructed \D\ candidate and at least one lepton
in the event with a momentum greater than 800\mevc~\cite{cm-frame}.
\DZ\ meson candidates are formed by \KM\PIP\ combinations requiring the
invariant mass to be consistent with the \DZ\ mass: 
$1.846 < m(K\pi) < 1.877 \gevcc$.
This asymmetric mass window is chosen to take into account resolution effects of
the detector. The selection is optimized to maximize the significance of the
selected sample.

\DZ\ candidates are combined with charged and neutral pions to form
\DS\ candidates. 
For \DSZ\ the \PIZ\ is reconstructed from a photon pair with an
invariant mass of $115 < m_{\gaga} < 150 \mevcc$. Those photon pairs are re-fitted
in a ``mass-constrained'' fit to match the nominal mass of the \PIZ.
\DS\ candidates are selected by their mass difference to the \DZ\ candidate:
$144 < m(\DZ\PIP) - m(\DZ) < 148\mevcc$ and
$140 < m(\DZ\PIZ) - m(\DZ) < 144 \mevcc$ for charged and neutral \DS, respectively.

\DP\ candidates are formed from \KM\PIP\PIP\ combinations with an invariant mass
of $1.854 < m(K\pi\pi) < 1.884 \gevcc$. The $\chi^2$ fit probability for the three tracks
to originate from a common vertex, $P_{\mbox{\scriptsize Vtx}}$, is required to be
$P_{\mbox{\scriptsize Vtx}}(K\pi\pi)>0.01$.

Candidates for \D\ and \DS\ are combined with charged pions to form \DSS\ candidates,
and finally paired with muons or electrons. The charge of the
lepton is required to match the charge of the kaon from the \D\ decay.

Part of the background is due to events where a \DSS\ is paired to a lepton from
the other $B$. Thus we require that the probability that the lepton and the pion emitted
by the \DSS\ originate from a common vertex exceeds 0.001, and that the angle between
the direction of flight of the \DSS\ and the lepton is more than 90 degrees.

A large fraction of the background events is due to $B\to D^*\ell\nu$ decays
where the \DS\ or its daughter \D\ is paired to a pion from the other $B$. 
To suppress this combinatorial background, we make use of the variable \COSBY\
described in the following.
The energy and momentum of the $B$ mesons from the \FourS\ decays are known from
incident beam energies. For correctly reconstructed \BDSSLN\ decays, where
the only missing particle is the neutrino, the decay kinematics can be calculated,
up to one angular quantity, from the four-momentum of the visible decay
products ($Y=\DSSL$).
The cosine of the angle between the direction of flight of the $B$ meson and its
visible decay product $Y$ is given by
\begin{equation}
\COSBY = 
  - \frac{2 E_B E_Y - m_B^2 -m_Y^2}{2 |\vec p_B| |\vec p_Y|}, \nonumber
\label{eqn:cosby}
\end{equation}
where $E$, $|\vec p|$ and $m$ are the energies, momenta and masses of the $B$ and
the $Y$, respectively.
If the $Y$ candidate is not from a correctly reconstructed \BDSSLN\ decay,
the quantity \COSBY\ no longer represents an angle, and can take any value.
We select candidates having $|\COSBY|\le 1$.

In case a \DS\ is reconstructed in the decay chain, a veto is applied against
decays $B\TO\DS\ell\nu$ by calculating the variable \COSBYP\ which
is defined as above, but the $Y$ system is redefined to contain only the \DS\ and
the lepton: $Y^{\prime}=\DS\ell$. Background events are rejected by the requirement
$\COSBYP<-1$ since signal events $\BDSSLN$ tend to have values less than $-1$.

To reduce combinatorial backgrounds in the decay chain \DSS\TO\D\PIM,
only the \DSSL\ candidate with $\tilde m_{\nu}^2$ closest to zero is selected, where
$\tilde m_{\nu}^2$ is the neutrino mass squared, calculated in the approximation $\vec p_B=0$:
$\tilde m_{\nu}^2 = m_B^2 + |\vec p_Y|^2 - 2 E_B E_Y$.
Events reconstructed in the \DSS\TO\DZ\PIM\ final state are rejected if the \DZ\ can be paired
with any charged pion to form a \DSP\ candidate as described above.

In about 2\% of the events more than one \DSSL\ candidate is selected
and if so all of them enter the analysis.

We determine the \DII\ signal yield in the channel $\DSS \to \D\PI$ and the
\DI\ and \DII\ signal yields in the channel $\DSS \to \DS\PI$ by a
binned $\chi^2$ fit to the $\dM=m(D^{(*)}\PIM)-m(\DZ)$ distributions.
To determine the individual contributions from \DI\ and \DII\ in the $\DS\PI$
final state, we make use of the helicity angle distribution of the \DS, $\vartheta_h$,
which is defined as the angle between the two pions emitted by the \DSS\ and the \DS\ in
the rest frame of the \DS.
For a \DS\ from a \DII\ this distribution varies as $\sin^2 \vartheta_h$,
whereas for \DI\ decays, the helicity angle is distributed
like $1+\ADI\cos^2\vartheta_h$,
where \ADI\ is a parameter which depends on the initial polarization of the \DI\
and a possible S wave contribution to the \DI\ decay.
To exploit this feature, we split the data for the two decay chains involving a
\DS\ into four subsamples, corresponding to four equal size bins in $|\cos \vartheta_h|$.

The resulting ten \dM\ distributions are fitted simultaneously
to determine 12 parameters describing the signal yields and distributions, and 22
parameters to adjust the background yields and shapes.
The mass differences for the signal events are described by Breit-Wigner functions.
There are four parameters giving the signal yields for the semileptonic decays 
involving the two narrow states, charged and neutral.
The masses of the states are also fitted, but are constrained to be equal for
charged and neutral states, giving two parameters.
Four additional parameters arise from the effective widths of the \DSS\ states, which
represent a convolution of the intrinsic widths and detector resolution effects.
The latter contributes approximately 2-3\mevcc, depending on the mode. 
The fit also determines the \DII\ branching ratio
$\BDII=\Gamma(\DII\TO\D\PIM)/\left(\Gamma(\DII\TO\D\PIM)+\Gamma(\DII\TO\DS\PIM)\right)$
and the \DI\ polarization amplitude \ADI.

Backgrounds are modeled by cubic functions in \dM.
The background shape in the \DS\PIM\ channel is found to be the same in all
helicity bins for each final state.
The fit thus has three shape parameters for each decay chain, while the number of
background events is determined independently in each bin.

The selection efficiency is deduced from a fit to the simulation. This fit
uses the same parametrization as the fit determining the signal yield
from data and is applied to the sum of the full background simulation
and for one signal decay chain at a time. For a given decay mode the efficiencies
are found to be the same for \DI\ and \DII, specifically:
$\epsilon(\DSP\PIM)= (6.89\pm 0.12)\%$,
$\epsilon(\DSZ\PIM)= (5.34\pm 0.12)\%$,
$\epsilon(\DP\PIM) = (12.88\pm 0.96)\%$ and
$\epsilon(\DZ\PIM) = (17.56\pm 0.70)\%$,
where the quoted uncertainties are the statistical uncertainties from the fit.
For the decays including a \DS\ the efficiency is multiplied by the probability for
a \DSS\ to decay with a value of $|\cos\vartheta_h|$ falling into a given bin.
This factor includes the theoretical distribution discussed above as well as
corrections for the different detector acceptances in the four helicity bins of
up to 10\%.
The total number of $B$ mesons in the data sample used for the present work is
$N_{\BB}=(236.0\pm2.6)\times 10^6$~\cite{bcount}.
For the charged and neutral $B$ mesons we assume
$\Gamma(\FourS\to\BpBm)/\Gamma(\FourS\to\BzBzb)=1.065\pm 0.026$~\cite{hfag}.

\begin{figure*}[t!]
\begin{center}
\resizebox{170mm}{!}{\includegraphics{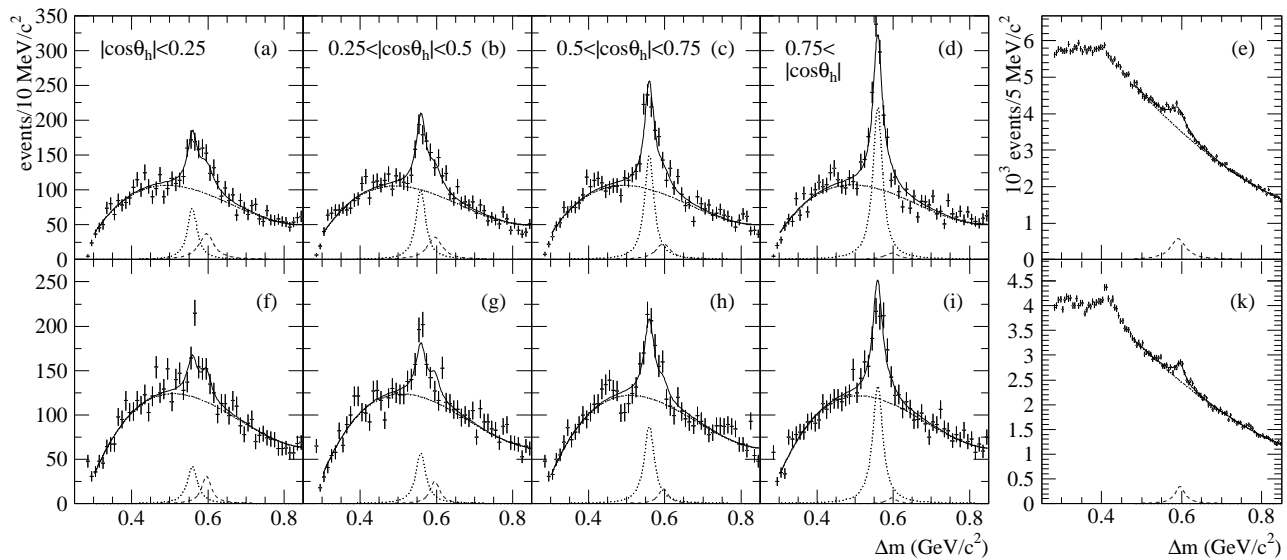}}
\caption{\label{fig:fit}
\dM-spectra for the selected data and the results of the fitted functions.
The solid line represents the complete fit function, dotted (\DI) and dashed (\DII) lines
for the signal and dash-dotted the for background.
(a) to (d) show the mode \DSSZ\TO\DSP\PIM\ with increasing values for $|\cos\vartheta_h|$,
(e) the mode \DSSZ\TO\DP\PIM. (f) to (i) show the corresponding bins in $|\cos\vartheta_h|$
for the mode \DSSP\TO\DSZ\PIP\ and (k) the mode \DSSP\TO\DZ\PIP.}
\end{center}
\end{figure*}

The fit procedure has been extensively validated.
The analysis procedure is tested on statistically independent MC simulated data
samples and was found to reproduce the
input signal parameters with a $\chi^2/n=12.66/12$, where $n$ is the number of
signal parameters.
Consistent fit results were also obtained when the data sample was separated into
subsamples representing specific data taking periods, separated by lepton species
or restricting it to certain decay modes,
using charged or neutral \DSS\ only, or combining the helicity bins.
The results of the fit are shown in Fig.~\ref{fig:fit}. As expected, the contribution
of the \DII\ vanishes for large values of $|\cos\vartheta_h|$ while the
contribution of the \DI\ is suppressed for $\cos\vartheta_h$ close to zero. The extracted
yields are given in Table \ref{tab:yields}.

\begin{table}[b!]
\begin{center}
\caption{\label{tab:yields}
Extracted yields for the four signal modes in the five relevant \dM-spectra.}
\begin{tabular}{ll@{$|$}rrrrr}
\hline \hline
mode & \multicolumn{2}{c}{$|\cos\vartheta_h|$} & 
\hspace*{3mm} \DIZ & \hspace*{3mm} \DIIZ & \hspace*{3mm} \DIP & \hspace*{3mm} \DIIP \\
\hline
\DS\PIP & $[0.00$&$0.25]$      &  344 &  273 & 212 &  152 \\
\DS\PIP & $[0.25$&$0.50]$       &  470 &  238 & 286 &  123 \\
\DS\PIP & $[0.50$&$0.75]$      &  699 &  170 & 439 &   83 \\
\DS\PIP & $[0.75$&$1.00]$         & 1027 &   67 & 668 &   31 \\
\D\PIP  & \multicolumn{2}{c}{} & $\cdots$ & 8414 & $\cdots$ & 3361 \\
\hline \hline
\end{tabular}
\end{center}
\end{table}

Systematic uncertainties have been analyzed and their impact on the fitted yields have
been estimated taking into account correlations between fit parameters.
Efficiencies for reconstructing and selecting the particles of the final state
are derived from Monte Carlo simulation. The simulation of the tracking and the
\PIZ\ reconstruction have been studied by comparing $\tau$ decays to one and three charged tracks
and with or without a neutral pion. Uncertainties introduced by the particle identification for
kaons and leptons are studied using control samples with high purities for the particles in
question. The impact of the finite statistics of the simulated signal events is deduced from the
fit error of the efficiency determination.

The uncertainty on the number of charged and neutral $B$ mesons in the data set is determined
as in~\cite{bcount,hfag} and the branching fractions of the decays of the \DS\ and the
\D\ are taken from~\cite{pdg}.

\begin{table}[b!]
\begin{center}
\caption{\label{tab:systematics}
Summary of systematic uncertainties of the
determination of the semileptonic branching fractions.}
\begin{tabular}{lcccc}
\hline \hline
Source & \multicolumn{4}{c}{$\Delta \BR (B\to\DSS\ell\nu)/ \BR (B\to\DSS\ell\nu) [\%]$}\\
& \hspace*{3mm}\DIZ\hspace*{3mm} & \hspace*{3mm}\DIIZ\hspace*{3mm} 
& \hspace*{3mm}\DIP\hspace*{3mm} & \hspace*{3mm}\DIIP\hspace*{3mm} \\
\hline
tracking                    & 1.76  & 1.39  &  1.03 &  1.14  \\
\PIZ\ efficiency            & 0.06  & 0.29  &  3.25 &  0.60  \\
particle identification     & 2.61  & 2.75  &  3.11 &  1.60  \\
MC statistics               & 1.80  & 5.61  &  2.50 &  3.32  \\
helicity correction         & 0.65  & 0.14  &  0.17 &  0.31  \\
number of $B$ mesons        & 2.68  & 2.68  &  2.68 &  2.68  \\
$\BR(\DSP\TO\DZ\PIP)$       & 0.76  & 0.19  &  0.04 &  0.10  \\
$\BR(\DSZ\TO\DZ\PIZ)$       & 0.11  & 0.45  &  5.07 &  0.93  \\
$\BR(\DZ\TO\KM\PIP)$        & 1.89  & 0.42  &  1.78 &  2.03  \\
$\BR(\DP\TO\KM\PIP\PIP)$    & 0.07  & 2.67  &  0.24 &  0.54  \\
signal modeling             & 2.11  & 4.75  &  3.21 &  1.95  \\
bkg. parametrization        & 1.93  & 1.68  &  3.20 &  2.71  \\
\hline
total                       & 5.76  & 9.03  &  9.16 &  6.17  \\
\hline \hline 
\end{tabular}
\end{center}
\end{table}

Uncertainties introduced by the physics model which was used to simulate the MC data
have been addressed by re-weighting the signal MC to an alternative decay model based on
HQET~\cite{LLSW}. The fit was repeated with efficiencies deduced from the
reweighted signal MC and the deviations in the results are taken as systematic uncertainties.
A possible influence of the background description has been tested by varying the
parametrizations. The backgrounds are alternatively described by a square root function,
$f(\dM)=\sqrt{\dM-m_0}$, where $m_0$ is the kinematic limit,
multiplied by either polynomials or exponentials in $\dM$.

Table \ref{tab:systematics} gives a summary of the various sources of
systematic uncertainty and their impact on the results.
Added in quadrature the total systematic uncertainties in the semileptonic
branching fractions are 6-10\%, depending on the \DSS\ type. 

In summary, we have measured the four branching fractions of $B$ mesons decaying
semileptonically into narrow \DSS\ states.
The \DSS\ decay rates are unknown, thus we can only determine the product branching fractions:
\begin{widetext}
\begin{eqnarray}
\BR (\Bp \to \DIZ  \ell^+ \nu_{\ell}) \times \BR(\DIZ\TO\DSP\PIM) 
&=& ( 2.97 \pm 0.17_{\mbox{\scriptsize stat}} \pm 0.17_{\mbox{\scriptsize syst}})\PERM, \nonumber \\
\BR (\Bp \to \DIIZ  \ell^+ \nu_{\ell})\times \BR(\DIIZ\TO\D^{(*)}{}^{+}\PIM) 
&=& ( 2.29 \pm 0.23_{\mbox{\scriptsize stat}} \pm 0.21_{\mbox{\scriptsize syst}})\PERM, \nonumber \\
\BR (\Bz \to \DIM  \ell^+ \nu_{\ell}) \times \BR(\DIM\TO\DSZ\PIM) 
&=& ( 2.78 \pm 0.24_{\mbox{\scriptsize stat}} \pm 0.25_{\mbox{\scriptsize syst}})\PERM, \nonumber \\
\BR (\Bz \to \DIIM  \ell^+ \nu_{\ell})\times \BR(\DIIM\TO\D^{(*)}{}^{0}\PIM) 
&=& ( 1.77 \pm 0.26_{\mbox{\scriptsize stat}} \pm 0.11_{\mbox{\scriptsize syst}})\PERM. \nonumber
\end{eqnarray}
\end{widetext}
We observe all modes with significance greater than $5\sigma$,
among them evidence of the \DIIM\ contribution to the decay $B\to\DS\pi\ell\nu$.
For modes already observed we find results in agreement with previous
measurements, but achieve
better precisions~\cite{DSS-CLEO,DSS-recoil,DSS-Z0}.

For the decays of the \DSS\ we measure the branching ratio
$\BDII = 0.62 \pm 0.03_{\mbox{\scriptsize stat}} \pm 0.02_{\mbox{\scriptsize syst}}$.
This ratio is in agreement with theoretical predictions~\cite{DSS-theory} and
previous measurements~\cite{pdg} but has a smaller uncertainty by a factor of about four.

For the \DI\ we determine the polarization parameter to be
$\ADI = 3.8 \pm 0.6_{\mbox{\scriptsize stat}} \pm 0.8_{\mbox{\scriptsize syst}}$. It is the
first measurement of the \DI\ polarization, within the uncertainties consistent with
unpolarized \DI\ decaying purely via D wave, which gives the prediction $\ADI=3$, but
violates HQS \cite{LLSW}.

\input acknow_PRL.tex

\end{document}

%% file: authors_jul2008.tex
%
\author{B.~Aubert}
\author{M.~Bona}
\author{Y.~Karyotakis}
\author{J.~P.~Lees}
\author{V.~Poireau}
\author{E.~Prencipe}
\author{X.~Prudent}
\author{V.~Tisserand}
\affiliation{Laboratoire de Physique des Particules, IN2P3/CNRS et Universit\'e de Savoie, F-74941 Annecy-Le-Vieux, France }
\author{J.~Garra~Tico}
\author{E.~Grauges}
\affiliation{Universitat de Barcelona, Facultat de Fisica, Departament ECM, E-08028 Barcelona, Spain }
% ...
\collaboration{The \babar\ Collaboration}
\noaffiliation

%% file: acknow_PRL.tex
We are grateful for the excellent luminosity and machine conditions
provided by our \pep2\ colleagues, 
and for the substantial dedicated effort from
the computing organizations that support \babar.
The collaborating institutions wish to thank 
SLAC for its support and kind hospitality. 
This work is supported by
DOE
and NSF (USA),
NSERC (Canada),
CEA and
CNRS-IN2P3
(France),
BMBF and DFG
(Germany),
INFN (Italy),
FOM (The Netherlands),
NFR (Norway),
MES (Russia),
MEC (Spain), and
STFC (United Kingdom). 
Individuals have received support from the
Marie Curie EIF (European Union) and
the A.~P.~Sloan Foundation.

%% file: paper.bbl
\begin{thebibliography}{99}
\bibitem{conjugate}
Throughout this Letter, whenever a mode is given, the charge
conjugate is also implied.
\bibitem{cm-frame}
Unless explicitly stated otherwise, all energies, momenta and angles are
measured in the \epem\ center of mass frame.
\end{thebibliography}
